\documentclass{ws-procs975x65}
\usepackage{subfigure}

\begin{document}

\title{COSMIC-RAY SPECTRUM AND COMPOSITION\\WITH THE ICECUBE OBSERVATORY}

\author{ALESSIO TAMBURRO$^*$ FOR THE ICECUBE COLLABORATION}

\address{Bartol Research Institute and Department of Physics and Astronomy\\ 
University of Delaware, Newark, DE 19716, USA\\
$^*$E-mail: tamburro@udel.edu}


\address{}

\begin{abstract}
This paper reports on recent results from measurements of energy spectrum and
nuclear composition of galactic cosmic rays
performed with the IceCube Observatory at the South Pole
in the energy range between about 300~TeV and 1~EeV. 
\end{abstract}


\bodymatter

\section{Introduction}\label{sec::intro}

\begin{figure}[b]
  \begin{minipage}[]{0.57\textwidth}
    \centering
 \subfigure[]{  \includegraphics[width=\textwidth]{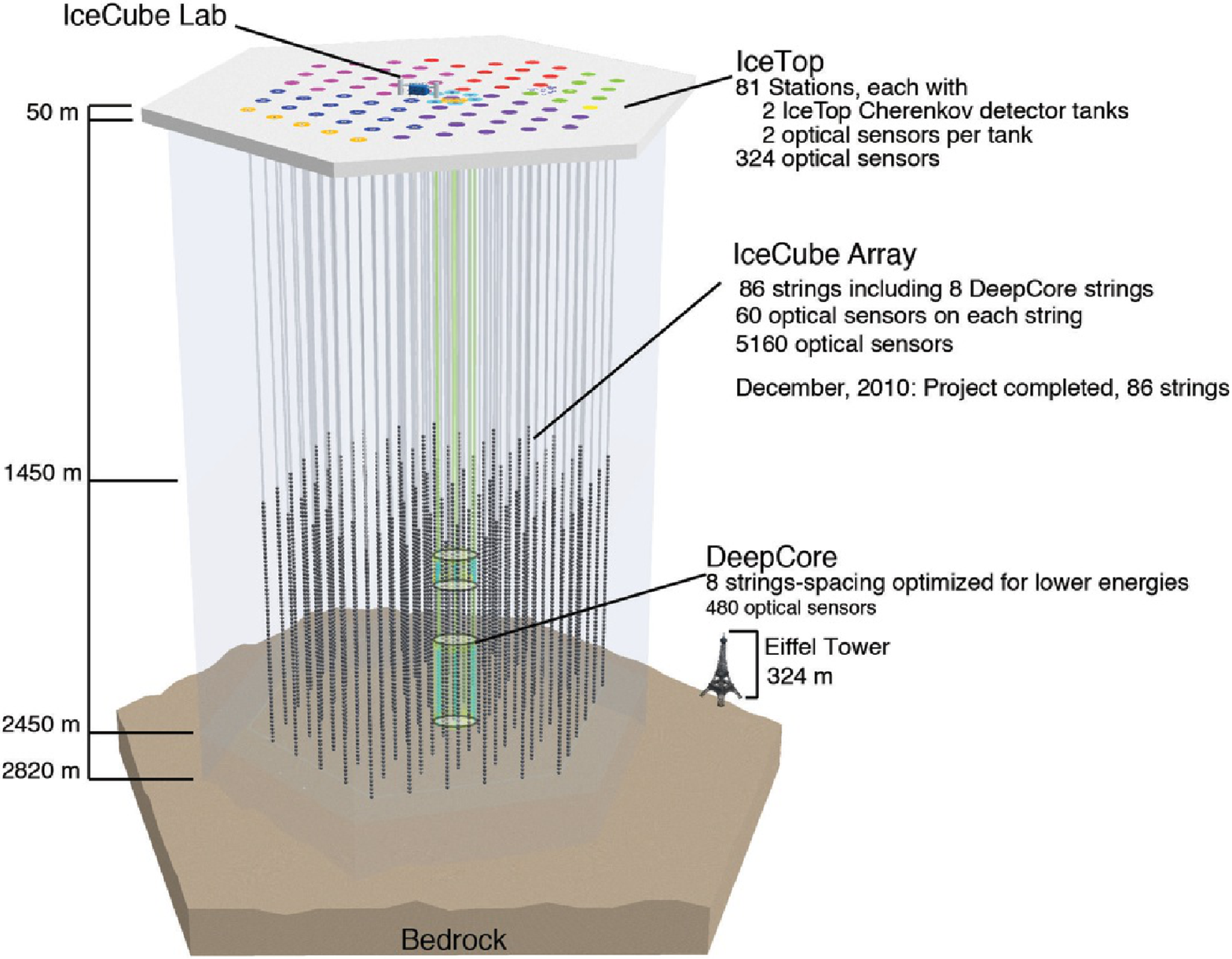}\protect\label{fig::icecube} }
 \end{minipage}\hfill
 \begin{minipage}[]{0.30\textwidth}
 \subfigure[]{\includegraphics[width=\textwidth]{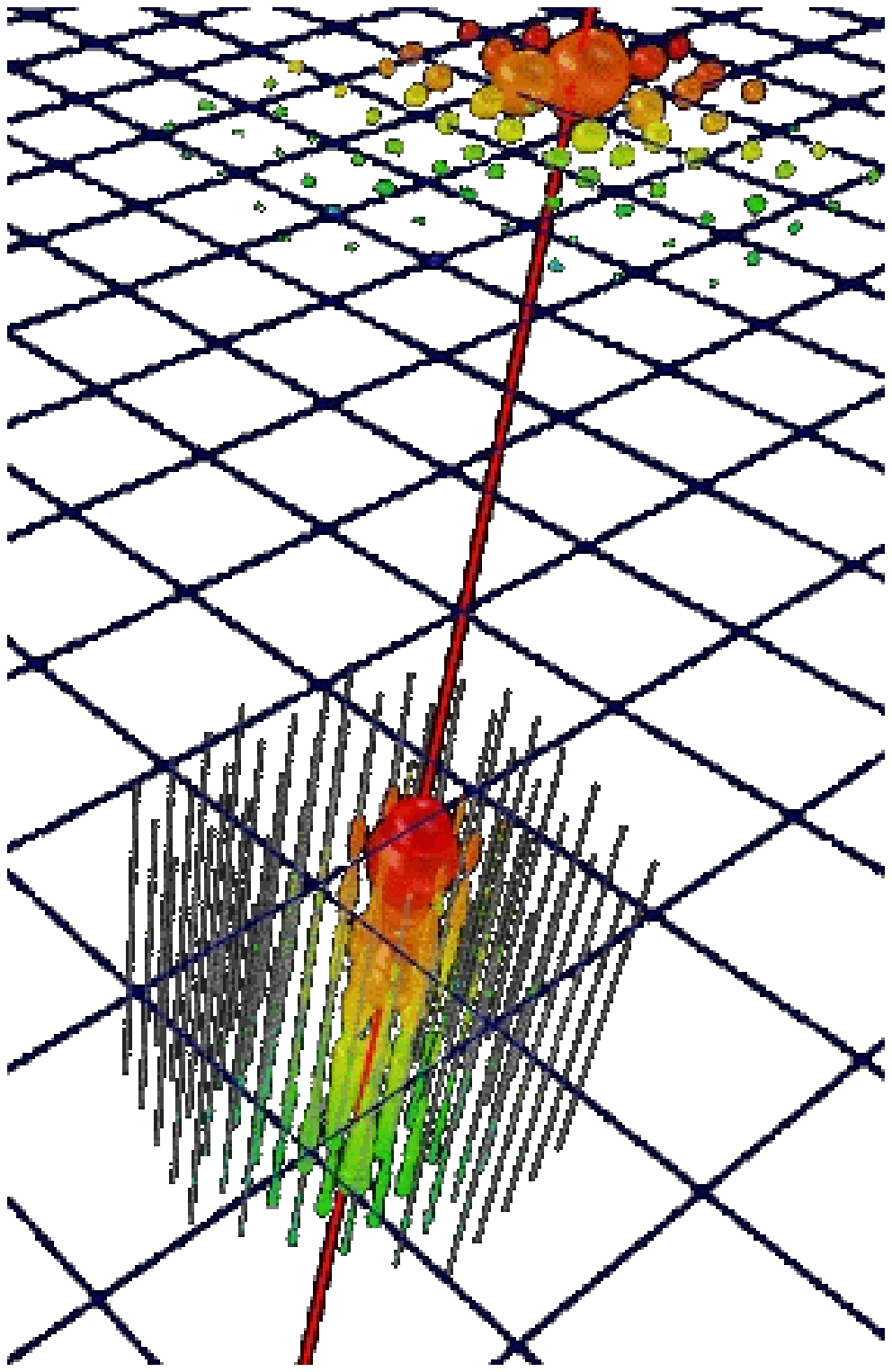}\protect\label{fig::coinc}}
 \end{minipage}
 \caption{
 (a) Sketch of the IceCube Observatory.
IceCube in its 2006-07 configuration is shown in red
and termed as IT26/IC22 (26 IceTop stations/22 in-ice strings). 
Other configurations are IT40/IC40 (2007-08) in green, 
IT59/IC59 (2008-09) in violet, IT73/IC79 (2009-10) in blue, and
IT81/IC86 (2010-11) in yellow.
 (b) Coincident event recorded in 2010 (IT73/IC79). 
 Colored spheres indicate DOM signals (the earliest ones being in red) with strength proportional to the radius.}
\end{figure}
Most galactic cosmic rays are believed to be accelerated 
in the blast waves of nearby supernova remnants, reaching
a maximum energy that scales with the charge of the nucleus.
The heaviest elements can gain up to about 10$^{18}$~eV.
The signatures of these sources are the gradual steepening
of the cosmic-ray flux at a few 10$^{15}$~eV, called the {\it knee},
and possibly more structures at higher energies.

The IceCube Observatory\cite{fyp} (Fig.~\ref{fig::icecube}) is a three-dimensional cosmic-ray air shower
detector measuring primary cosmic particles
with energy $E$ between about 3~$\cdot 10^{14}$~eV 
and $10^{18}$~eV. The observatory has been in operation since May 2011 with 86 strings
between 1.45 and 2.45 km below the surface.
Each string carries 60 digital optical modules\cite{DOM}  (DOMs)
which include photomultipliers and readout electronics. 
Near the top of each string (2835~m altitude, atmospheric depth 
of about 680~g/cm$^2$), 81 pairs (stations) of cylindrical Cherenkov tanks
cover about 1~km$^2$ and are part of the surface component of IceCube, named IceTop\cite{IT}.
Each tank contains two standard IceCube DOMs	
and samples low-energy photons, electrons, and muons from air showers.
The deep detectors measure the signal of penetrating muons
(more than about 500~GeV) from the early stage of shower development.

In Fig.~\ref{fig::coinc} an event seen in coincidence by both the surface 
and the deep detectors (coincident event) is shown.
The effective area of IceCube for coincident events is
A$\approx$0.15~km$^2$sr (0.4~km$^2$sr for IceTop alone). 
The maximum energy above which the intensity is too low to obtain
enough events for analysis is about 1~EeV.

\section{Primary Energy Spectrum and Nuclear Composition}\label{sec::results}

\begin{figure}[t]
\centering
\includegraphics[width=8.0cm]{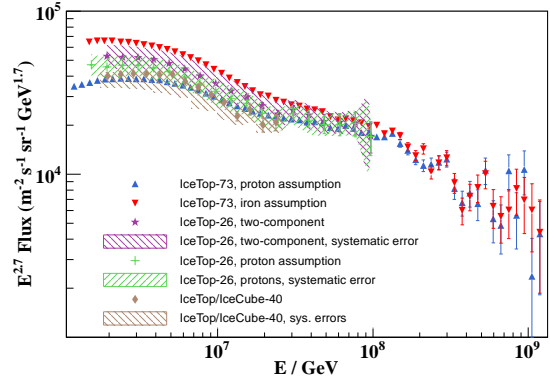}
\caption{Energy spectra obtained with data of IceTop running in different configurations 
as labeled.
\protect\label{fig::spectraIT}}
\end{figure}
The position of IceTop at the high altitude of the South Pole
makes it possible to sample secondaries near the shower maximum,
thus reducing the effects of fluctuations.
The energy resolution is 0.1 or less in units of $\log_{10}($E/GeV$)$
above about 1~PeV (0.05 above 10~PeV).
The first analysis to determine the all-particle energy spectrum\cite{Abbasi:2012wn}
(Fig.~\ref{fig::spectraIT}) was based on data of IT26 (area of 0.094~km$^2$, June to October 2007).
Assuming mixed composition (H and Fe only), 
the knee is measured at about 4.3~PeV and
the spectral index above the knee is about -3.1. An indication of a flattening of
the spectrum above 22~PeV is also observed with a spectral index changing to about -2.9.
A preliminary measurement of the IT73 spectrum was 
obtained by analyzing 11 months of data (June 2009 to 
May 2010). The statistics is nearly 4$\cdot$10$^7$ events. Of these events,
about 200 are found above about 200~PeV.

\begin{figure}[t]
\centering
\includegraphics[width=8.0cm]{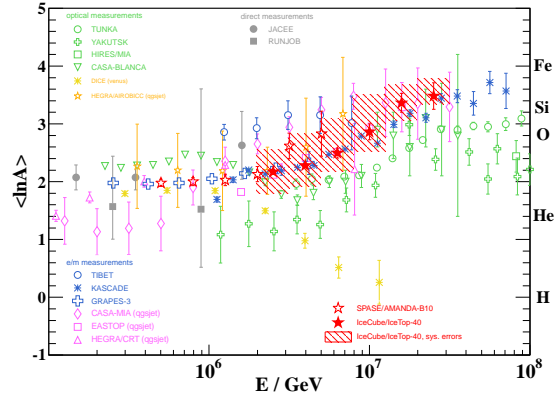}
\caption{The mean logarithmic mass $<lnA>$ vs primary energy adapted from 
Ref.\citen{IceCube:2012vv}. The result indicates a strong increase in mass through the knee
although the systematic uncertainties can greatly affect the measured composition in terms
of absolute value of $<lnA>$.}
\protect\label{fig::massc}
\end{figure}

A measurement of the primary mass composition (Fig.~\ref{fig::massc})
was performed with one month of data\cite{IceCube:2012vv} of IT40/IC40.
A neural network was trained with Monte Carlo simulations
of 5 primaries (H, He, O, Si, Fe).
Measurements of $e$/$\gamma$
component of air showers at the surface and $\mu$ component 
in the ice are used to ``teach'' the network how
to find the best fit to $E$ and mass. 

\section{Conclusions}\label{sec::conc}

The IceCube Observatory is currently taking data in the second year
after its completion. Based on data of the detector in earlier stages of its deployment,
events seen in coincidence by both the surface component
and in-ice detectors have been analyzed to 
investigate the mass composition of cosmic rays.
IceTop event analysis resulted in measurements
of the cosmic-ray energy spectrum. 
The results are in line with those of other experiments
but IceCube has the potential to yield high precision for energy and mass spectrum
from below the knee to about 1 EeV.

\section*{Acknowledgments}

I am grateful to T. Gaisser, H. Kolanoski, and T. Stanev
for  discussions. This research is supported in part by the
U.S. National Science Foundation Grant NSF-ANT-0856253.

\end{document}